# Handling errors in four-dimensional variational data assimilation by balancing the degrees of freedom and the model constraints: A new approach

Xiangjun Tian[1,4,*], Hongqin Zhang[2], Zhe Jin[3], Min Zhao[1], Yilong Wang[1],

Yinhai Luo[4], Ziqing Zhang[4], Yanyan Tan[1,4]

[1]State Key Laboratory of Tibetan Plateau Earth System, Environment and Resources (TPESER), Institute of Tibetan Plateau Research, Chinese Academy of Sciences, Beijing, 100101, China

[2]Institute of Atmospheric Physics, Chinese Academy of Sciences, 100029, Beijing, China

[3]Sino-French Institute for Earth System Science, College of Urban and Environmental Sciences, Peking University, Beijing 100871, China

[4]University of the Chinese Academy of Sciences, Beijing 100049, China

* Corresponding author. Xiangjun Tian (tianxj@itpcas.ac.cn)

**Abstract** For many years, strongly and weakly constrained approaches were the only options to deal with errors in four-dimensional variational data assimilation (4DVar), with the aim of balancing the degrees of freedom and model constraints. Strong model constraints were imposed to reduce the degrees of freedom encountered when optimizing the strongly constrained 4DVar problem, and it was assumed that the models were perfect. The weakly constrained approach sought to distinguish initial errors from model errors, and to correct them separately using weak model constraints. Our proposed i4DVar[*] method exploits the hidden mechanism that corrects initial and model errors simultaneously in the strongly constrained 4DVar. The i4DVar[*] method divides the assimilation window into several sub-windows, each of which has a unique integral and flow-dependent correction term to simultaneously handle the initial and model errors over a relatively short period. To overcome the high degrees of freedom of the weakly constrained 4DVar, for the first time we use ensemble simulations not only to solve the 4DVar optimization problem, but also to formulate this method. Thus, the i4DVar* problem is solvable even if there are many degrees of freedom. We experimentally show that i4DVar* provides superior performance with much lower computational costs than existing methods, and is simple to implement.

**Plain Language Summary** The strongly 4DVar ignores the model error and only corrects the initial condition error at the expense of reduced accuracy; while the weakly 4DVar accounts for both the initial and model errors but corrects them separately, which increases computational costs and uncertainty. Remarkably, the strongly constrained 4DVar has a hidden mechanism that can correct all the system-related errors together as a whole at the analysis time. In this study, this hidden mechanism is further extended by the proposed i4DVar* method to suppresses both the initial and model error evolution over a relatively short period by adding flow-dependent correction terms at defined time steps of identical intervals $\tau$ in the entire assimilation window. This novel approach led to greatly improved assimilation and forecast results than existing methods with minimum extra costs in all our numerical experiments.

**Key Points**

- Currently, strongly and weakly constrained approaches were the only options to deal with errors in 4DVar.
- i4DVar$^*$ divides the assimilation window into several sub-windows to correct all the system-related errors together flow-dependently.
- For the first time we use ensemble simulations not only to solve the 4DVar optimization problem, but also to formulate this method.

1. **Introduction**

Data assimilation (DA), which is accomplished using evolving four-dimensional variational data assimilation (4DVar) approaches (Le Dimet & Talagrand, 1986; Lewis & Derber, 1985), is attracting enormous attention in the present era of big data (Argaud et al., 2009; Bouittier & Kelly, 2001; Clayton et al., 2013; Elbern et al., 2000; Fisher & Lary, 1995; Gauthier & Thépaut, 2001; Kaminski et al., 2013; Lellouche et al., 2000; Lewis et al., 2012; Luo et al., 2011; McNally, 2009; Mitchell et al., 2004; Rabier et al., 2000). DA incorporates enormous numbers of observations into numerical simulations, corrects errors, and improves forecasting. 4DVar methods are optimal for initializing the numerical weather predictions (NWPs) of several major operational entities (e.g., the European Centre for Medium-Range Weather Forecasts [ECMWF] and the "weather hubs" of France, UK, Japan, and China).

4DVar methods aim to identify the optimal model states $x_i$ at each time step $t_i$ ($i = 1, \cdots, N_{\Delta t}$, where $N_{\Delta t} = \frac{t_S - t_0}{\Delta t}$ and $\Delta t$ is the model time step) by assimilating observational vectors $\mathbf{y}_{o,k} \in \mathbb{R}^{m_{y,k}}$ at different time levels $t_k$ over an assimilation window $[t_0, t_S]$. In other words, the analytical increments $x_i'$ of the background ("first-guess") model states $x_{b,i}$ are the optimized variables of degrees of freedom $\mathbf{D}_{\text{OF}}$ about $N_{\Delta t} \times N_v \times n_{\text{dim}}$ ($n_{\text{dim}} = n_x \times n_y \times n_z$, $m_x = N_v \times n_{\text{dim}}$), where $N_v$ is the number of analytical variables and $n_x, n_y, n_z$ are the numbers of model grids in the $x$, $y$, and $z$ directions, respectively. Usually, $\mathbf{D}_{\text{OF}} \gg m_y$ (where $m_y = \sum m_{y,k}$ is the dimension of the observational vectors); at this moment, the forecast model (from $t_0$ to $t_k$) $\mathbf{x}_k = M_{t_0 \to t_k}(\mathbf{x}_0)$ has no constraints during 4DVar optimization and underdetermination is possible.

To solve this four-dimensional (4D) optimization problem, it has been necessary to impose model constraints; the strongly constrained 4DVar (s4DVar) uses the analytical increment $\mathbf{x}' \in \mathbb{R}^{m_x}$ ($= \mathbf{x}_0 - \mathbf{x}_b$) under the initial conditions, such that $\mathbf{x}'$ minimizes the cost function:

$$J(\mathbf{x}') = \frac{1}{2}(\mathbf{x}')^T \mathbf{B}(\mathbf{x}') + \frac{1}{2}\sum_{k=0}^{S}[H_k(\mathbf{x}_k) - \mathbf{y}_{o,k}]^T \mathbf{R}_k^{-1}[H_k(\mathbf{x}_k) - \mathbf{y}_{o,k}] \quad (1)$$

under the constraint $\mathbf{x}_k = M_{t_0 \to t_k}(\mathbf{x}_0)$. Here, the superscript T indicates the matrix transpose, while $\mathbf{x}_b$ is the background field at the initial time $t_0$, $S+1$ is the total number of observations in the assimilation window, $H_k$ is the observational operator, and matrices $\mathbf{B} \in \mathbb{R}^{m_x \times m_x}$ and $\mathbf{R}_k \in \mathbb{R}^{m_{y,k} \times m_{y,k}}$ are the background and observation error covariances, respectively. It is apparent that the optimized variable of (1) is simply the analytical increment $\mathbf{x}'$ at $t_0$, the dimensions of which are only $N_v \times n_{\text{dim}}$, i.e., much lower than the original $\mathbf{D}_{OF} = N_{\Delta t} \times N_v \times n_{\dim}$. The adjoint (Le Dimet & Talagrand, 1986; Lewis & Derber, 1985; Courtier & Talagrand, 1987) and incremental (Courtier et al., 1994) approaches made 4DVar feasible, such that it was subsequently applied in many major operational NWP centers. Traditional (i.e., strongly constrained) 4DVar assumes that the forecast model $\mathbf{x}_k = M_{t_0 \to t_k}(\mathbf{x}_0)$ precisely describes the underlying natural system. In other words, model error is not considered; it is implicit that any such error is minor compared to those under the initial conditions. However, the assumption of a perfect model is in fact not required (Tian et al., 2021). Traditional 4DVar is formulated as a constrained optimization problem that cannot show, and does not consider, whether the forecast is a perfect representation of the natural system. However, model error is often non-negligible given errors in the discretization of continuous fields, parameter uncertainties, boundary conditions, and round-offs. Given that model errors are inevitable, and that 4DVar operates successfully in major NWP centers, strongly constrained 4DVar has a hidden mechanism that simultaneously corrects initial and model errors. From this viewpoint, the analytical increment of traditional 4DVar is essentially an integral correction term focused on the initial time point. This indiscriminately counterbalances both the initial and model errors over the entire assimilation window (Tian et al., 2021).

Over time, researchers began to realize that model errors cannot be ignored, and then sought to distinguish initial and model errors and correct them separately in weakly constrained 4DVar approaches (w4DVar, Shaw & Daescu, 2017; Trémolet, 2006, 2007). Such approaches aimed to identify the analytical increments of the initial conditions, $\mathbf{x}'$, and the model errors $\varepsilon_k \in \mathbb{R}^{m_x}$, such that $\mathbf{x}'$ and $\varepsilon_k$ together minimized the following cost function:

$$J(\mathbf{x}', \varepsilon_1, \cdots, \varepsilon_S) = \frac{1}{2}(\mathbf{x}')^T \mathbf{B}(\mathbf{x}') + \frac{1}{2}\sum_{k=0}^{S}[H_k(\mathbf{x}_k) - \mathbf{y}_{o,k}]^T \mathbf{R}_k^{-1}[H_k(\mathbf{x}_k) - \mathbf{y}_{o,k}] \\ + \frac{1}{2}\sum_{k=1}^{S}(\varepsilon_k)^T \mathbf{Q}_k^{-1}(\varepsilon_k) \qquad (2)$$

subject to the state (constraint) equations $\mathbf{x}_k = M_{t_{k-1} \to t_k}(\mathbf{x}_{k-1}) + \varepsilon_k$, where $\varepsilon_k$ is the model error at $t_k$ and $\mathbf{Q}_k \in \mathbb{R}^{m_x \times m_x}$ are the model error covariance matrices. Although $\varepsilon_k$ differs from the analytical increments $\mathbf{x}'_k$ discussed above, the weakly constrained 4DVar (2) with many degrees of freedom [ $\mathbf{D}_{\text{OF}} = (S+1) \times N_v \times n_{\text{dim}}$ ] remains underdetermined if additional information is lacking. To address this issue and reduce computational costs, various simplifications have been used to specify the model errors of the weakly constrained 4DVar approaches (Shaw & Daescu, 2017; Griffith & Nichols, 2000). However, new issues arise; for example, it is very difficult and/or expensive to determine parameters that well-represent model error covariance (Shaw & Daescu, 2017). Such simplifications are essential when introducing additional information to constrain the optimization problem (2) when there are many degrees of freedom; an accurate solution is then possible. Both scale separation and new model error diagnostic techniques (Laloyaux et al., 2020a, 2020b) were used to overcome the two challenges (i.e., higher computational costs and uncertainty); a weakly constrained 4DVar was recently implemented at ECMWF. This probably represents the best effort to date to deal with model error during DA. However, this may not be the only way to deal with model error. Also, the ECMWF focused on the Integrated Forecast System (IFS) model (Laloyaux et al., 2020a); it is not clear whether the ECMWF error "fix" be applied to other models. Finally, a question arises as to whether the hidden mechanism of strongly constrained 4DVar that simultaneously and indiscriminately corrects both initial and model errors has practical importance or developmental significance.

The answer to this question is "yes". The integral-correcting 4DVar (i4DVar, Tian et al., 2021) extended the strategy used by the strongly constrained 4DVar to correct initial and model errors by introducing an averaged penalization term in the cost function, which corrected errors at defined time steps of identical intervals $\tau$, which is equivalent to dividing the assimilation window into several sub-windows (Tian et al., 2021). Noticeably, the averaged integral correction/analytical increment $\mathbf{x}'$ is added at the opening times of the sub-windows rather than at the opening time of the entire window. However, the optimized variable remains $\mathbf{x}'$ and the degrees of freedom are still $N_v \times n_{dim}$. Compared to the strongly constrained 4DVar, there is no change in the degrees of freedom. Thus, integral correction/analytical increment refers simply to the use of an averaged correction term to counterbalance both the initial and model errors, and to correct error evolution at selected time steps (Tian et al., 2021). Such an averaged/uniform correction term is clearly flow-independent over the entire assimilation window.

As the correction terms (i.e., $\mathbf{x}_k'$) tend to change according to the sub-windows, the degrees of freedom become $\mathbf{D}_{OF} = N_\tau \times N_v \times n_{dim}$ ($N_\tau = \frac{t_S - t_0}{\tau}$); thus, we have come "full circle". As discussed above, efforts to make weakly constrained 4DVars solvable led to simplifications (similar to those of parameterization schemes) to reduce the degrees of freedom. How do the properties of our i4DVar, which corrects the initial and model errors simultaneously, develop over time? Can i4DVar be further developed to truly provide a new method for dealing with model error by balancing the degrees of freedom and model constraints?

We developed an enhanced integral correcting 4DVar (i4DVar$^*$) that corrects errors by balancing the degrees of freedom and model constraints, thus improving 4DVar analyses and predictions (see Table 1 for comparisons between the s4DVar, w4DVar, i4DVar and i4DVar$^*$ methods). We show that the existing approaches balance the degrees of freedom and model constraints in a different way. To deal with the many degrees of freedom of the weakly constrained 4DVar, for the first time we used ensemble simulations not only to solve the optimization problem, but also to formulate this method. Finally, we performed numerical experiments comparing our i4DVar* with the strongly constrained 4DVar and the integral correcting one.

## 2. Materials and Methods
### 2.1. The enhanced i4DVar

Inspired by the i4DVar method (Tian et al., 2021), we created an enhanced integral-correcting 4DVar (i4DVar$^*$) via a 4D integral correction function $\mathbf{x}' = \left( \mathbf{x}_{t_0}'^T, \ \mathbf{x}_{t_1}'^T, \ \cdots, \ \mathbf{x}_{t_{N_\tau-1}}'^T \right)^T$ that minimizes the following cost function:

$$J(\mathbf{x}') = \frac{1}{2}(\mathbf{x}')^T \mathbf{Q}_b^{-1}(\mathbf{x}') + \frac{1}{2}\sum_{i=1}^{N_\tau}\left\{\sum_{k=t_{i-1}}^{t_i}\left[L_k'(\mathbf{x}_{t_{i-1}}') - \mathbf{y}_{o,i-1}^{k'}\right]^T \mathbf{R}_{i-1,k}^{-1}\left[L_k'(\mathbf{x}_{t_{i-1}}') - \mathbf{y}_{o,i-1}^{k'}\right]\right\}, \quad (3)$$

where $t_i = i \times \tau$ ($i = 1, \cdots, N_\tau$),

$$L_{k,i-1}'(\mathbf{x}_{t_{i-1}}') = L_{k,i-1}(\mathbf{x}_{b,i-1} + \mathbf{x}_{t_{i-1}}') - L_{k,i-1}(\mathbf{x}_{b,i-1}), \quad (4)$$

$$\mathbf{y}_{o,i-1}^{k'} = \mathbf{y}_{o,i-1}^k - L_{k,i-1}(\mathbf{x}_{b,i-1}), \quad (5)$$

$$L_{k,i-1} = H_k M_{t_{i-1} \to t_k}, \quad (6)$$

and:

$$\mathbf{x}_{k,i-1} = M_{t_{i-1} \to t_k}(\mathbf{x}_{b,i-1} + \mathbf{x}_{t_{i-1}}') \text{ in } [t_{i-1}, t_i]. \quad (7)$$

The background conditions $\mathbf{x}_{b,i-1}$ ($i = 1, \cdots, N_\tau$ and $\mathbf{x}_{b,0} = \mathbf{x}_b$) at the opening times of the sub-windows are as follows:

$$\mathbf{x}_{b,i-1} = M_{t_0 \to t_{i-1}}(\mathbf{x}_{b,0}), \quad (8)$$

where $\mathbf{Q}_b \in \mathbb{R}^{(m_x \times N_\tau) \times (m_x \times N_\tau)}$ is the integral background error covariance (Tian et al., 2021).

If the usual adjoint-based approach (Le Dimet & Talagrand, 1986; Lewis & Derber, 1985; Courtier & Talagrand, 1987) is utilized, it is impossible to solve eqs. (3–8) as there are too many degrees of freedom. To overcome this, we incorporate ensemble simulations as follows. First, we prepare model perturbations (MPs) $\mathbf{P}_{x,0} = (\mathbf{x}_{0,1}', \cdots, \mathbf{x}_{0,N}')$ at the opening time of the DA window $[t_0, t_S]$ (Tian et al., 2018).

Next, we perform ensemble simulations using the forecast model $M_{t_0 \to t_k}(\cdot)$ as follows:

$$\mathbf{x}_{i-1,j} = M_{t_0 \to t_{i-1}}(\mathbf{x}_{b,0} + \mathbf{x}_{0,j}'), \quad (9)$$

where $j = 1, \cdots, N$ ($N$ is the ensemble size). In the third step, we define MPs $\mathbf{P}_{x,i-1}$ at the opening times of the sub-windows, and thus all MPs $\mathbf{P}_x$, as:

$$\mathbf{x}_{i-1,j}' = M_{t_0 \to t_{i-1}}(\mathbf{x}_{b,0} + \mathbf{x}_{0,j}') - \mathbf{x}_{b,i-1}, \quad (10)$$

$$\mathbf{P}_{x,i-1} = (\mathbf{x}_{i-1,1}', \cdots, \mathbf{x}_{i-1,N}'), \quad (11)$$

and

$$\mathbf{P}_x = \begin{pmatrix} \mathbf{P}_{x,0}^T, & \mathbf{P}_{x,1}^T, & \cdots, & \mathbf{P}_{x,N_\tau-1}^T \end{pmatrix}^T \tag{12}$$

Finally, we assume that the solution space of $\mathbf{x}' = \begin{pmatrix} \mathbf{x}'_{t_0}{}^T, & \mathbf{x}'_{t_1}{}^T, & \cdots, & \mathbf{x}'_{t_{N_\tau-1}}{}^T \end{pmatrix}^T$ is the linear space $\Omega(\mathbf{P}_x)$ spanned by $\mathbf{P}_x$ [i.e., $\mathbf{x}' \in \Omega(\mathbf{P}_x)$]. Thus, the "integral" correction $\mathbf{x}'$ can be expressed as linear combinations of the MPs, as follows:

$$\mathbf{x}' = \mathbf{P}_x \boldsymbol{\beta}. \tag{13}$$

The integral background error covariance $\mathbf{Q}_b$ can be approximated as follows (Tian et al., 2021):

$$\mathbf{Q}_b = \frac{(\mathbf{P}_x)(\mathbf{P}_x)^T}{N-1}. \tag{14}$$

Note that, on the one hand, eqs. (7–8) impose model constraints on the 4DVar optimization problem with many degrees of freedom ($N_\tau \times N_v \times n_{\dim}$). On the other hand, Eqs. (9–11) yield a linear solution space $\Omega(\mathbf{P}_x)$. Thus, eqs. (3–14) render the i4DVar* problem solvable. See figure 1 for the schematic diagram of i4DVar* vs. 4DVar.

Substituting eq. (13) and the ensemble error covariance $\mathbf{Q}_b = \frac{(\mathbf{P}_x)(\mathbf{P}_x)^T}{N-1}$ (eq. 14) into eqs. (3-9) and expressing the cost function in terms of the new control variable $\boldsymbol{\beta}$ yield

$$\begin{aligned} J(\mathbf{x}') &= (N-1) \cdot \boldsymbol{\beta}^T \boldsymbol{\beta} + \\ &\quad \frac{1}{2} \sum_{i=1}^{N_\tau} \left\{ \sum_{k=t_{i-1}}^{t_i} \left[ L'_{k,i-1}(\mathbf{P}_{x,i-1}\boldsymbol{\beta}) - \mathbf{y}^{k'}_{o,i-1} \right]^T \mathbf{R}^{-1}_{i-1,k} \left[ L'_{k,i-1}(\mathbf{P}_{x,i-1}\boldsymbol{\beta}) - \mathbf{y}^{k'}_{o,i-1} \right] \right\} \\ &\approx (N-1) \cdot \boldsymbol{\beta}^T \boldsymbol{\beta} + \\ &\quad \frac{1}{2} \sum_{i=1}^{N_\tau} \left\{ \sum_{k=t_{i-1}}^{t_i} \left[ \mathbf{P}^k_{y,i-1}\boldsymbol{\beta} - \mathbf{y}^{k'}_{o,i-1} \right]^T \mathbf{R}^{-1}_{i-1,k} \left[ \mathbf{P}^k_{y,i-1} - \mathbf{y}^{k'}_{o,i-1} \right] \right\} \end{aligned}, \tag{15}$$

where

$$\mathbf{P}^k_{y,i-1} = \begin{pmatrix} \mathbf{y}^{k'}_{i-1,1}, \cdots, \mathbf{y}^{k'}_{i-1,N} \end{pmatrix} \tag{16}$$

and

$$\mathbf{y}^{k'}_{i-1,j} = L'_{k,i-1}(\mathbf{x}'_{i-1,j}) \tag{17}$$

Similarly, after a series of mathematical transformations similar to those in formulating ensemble nonlinear least squares-based 4DVar (NLS-4DVar, Tian et al., 2018), eq. (15) can be also transformed into a non-linear least squares formulation, which is solved by the Gauss–Newton iteration scheme as follows (Tian et al., 2018):

$$\begin{aligned}\boldsymbol{\beta}^l = \boldsymbol{\beta}^{l-1} &- \left[(N-1)\mathbf{I}_{N\times N} + (\mathbf{P}_y)^T \mathbf{R}^{-1}(\mathbf{P}_y)\right]^{-1} \\ &\times \left\{(\mathbf{P}_y)^T \mathbf{R}^{-1}\left[L'(\mathbf{P}_x\boldsymbol{\beta}^{l-1}) - \mathbf{y}'_o\right] + (N-1)\boldsymbol{\beta}^{l-1}\right\}\end{aligned} \tag{18}$$

for $l = 1, \cdots, l_{\max}$, where $l_{\max}$ is the maximum iteration number,

$$\mathbf{P}_y = \left((\mathbf{P}^{t_0}_{y,0})^T \cdots (\mathbf{P}^{t_1}_{y,0})^T \cdots (\mathbf{P}^{t_{N_\tau-1}}_{y,N_\tau-1})^T \cdots (\mathbf{P}^{t_{N_\tau}}_{y,N_\tau-1})^T\right)^T, \tag{19}$$

$$\mathbf{y}'_o = \left((\mathbf{y}^{t_0'}_{o,0})^T \cdots (\mathbf{y}^{t_1'}_{o,0})^T \cdots (\mathbf{y}^{t_{N_\tau-1}'}_{o,N_\tau-1})^T \cdots (\mathbf{y}^{t_{N_\tau}'}_{o,N_\tau-1})^T\right)^T, \tag{20}$$

$$\mathbf{R} = \begin{pmatrix} \mathbf{R}_{0,t_0} & 0 & 0 & 0 & 0 & \cdots & 0 \\ 0 & \ddots & 0 & 0 & 0 & \ddots & 0 \\ 0 & 0 & \mathbf{R}_{0,t_1} & 0 & \ddots & \ddots & 0 \\ 0 & 0 & 0 & \ddots & 0 & \ddots & 0 \\ 0 & 0 & 0 & 0 & \mathbf{R}_{N_\tau-1,t_{N_\tau-1}} & \ddots & \vdots \\ \vdots & \ddots & \ddots & \ddots & \ddots & \ddots & 0 \\ 0 & \cdots & 0 & 0 & 0 & 0 & \mathbf{R}_{N_\tau-1,t_{N_\tau}} \end{pmatrix}, \tag{21}$$

and

$$L'(\mathbf{x}') = \begin{pmatrix} L'_{k,0}(\mathbf{x}'_{t_0}) \\ L'_{k,1}(\mathbf{x}'_{t_1}) \\ \vdots \\ L'_{k,N_\tau-1}(\mathbf{x}'_{t_{N_\tau-1}}) \end{pmatrix}. \tag{22}$$

To filter out the spurious long-range correlations resulting from the finite ensemble number $N$, we follow (Tian et al., 2018) to localize eq. (18) as follows:

$$\boldsymbol{\beta}^l_\rho = \boldsymbol{\beta}^{l-1}_\rho + (\boldsymbol{\rho}_y <e> \mathbf{P}^*_y)^T L'(\mathbf{x}^{',l-1}) + (\boldsymbol{\rho}_y <e> \mathbf{P}^{\#}_y)^T \mathbf{R}^{-1}\left[\mathbf{y}'_o - L'(\mathbf{x}^{',l-1})\right], \tag{23}$$

and

$$\mathbf{x}'^{,l} = \left(\boldsymbol{\rho}_x <e> \mathbf{P}_x\right)\boldsymbol{\beta}_\rho^l = \left(\boldsymbol{\rho}_x \circ \mathbf{P}_{x,1}^*, \cdots, \boldsymbol{\rho}_x \circ \mathbf{P}_{x,N}^*\right)\boldsymbol{\beta}_\rho^l \qquad (24)$$

where

$$\mathbf{P}_y^* = -(N-1)\mathbf{P}_y\left[\left(\mathbf{P}_y\right)^T\left(\mathbf{P}_y\right)\right]^{-1}\left[(N-1)\mathbf{I}_{N\times N} + \left(\mathbf{P}_y\right)^T\mathbf{R}^{-1}\left(\mathbf{P}_y\right)\right]^{-1}, \qquad (25)$$

$$\mathbf{P}_y^* = \left(\mathbf{P}_y\right)\left[(N-1)\mathbf{I}_{N\times N} + \left(\mathbf{P}_y\right)^T\mathbf{R}^{-1}\left(\mathbf{P}_y\right)\right]^{-1} \qquad (26)$$

$\boldsymbol{\rho}_x = \mathbb{R}^{m_x \times r}$, $\boldsymbol{\rho}_x\boldsymbol{\rho}_x^T = \mathbf{C} \in \mathbb{R}^{m_x \times m_x}$, $\mathbf{C}$ is the spatial correlation matrix computed through

$$\mathbf{C}(i,j) = \mathbf{C}_0\left(\frac{d_{i,j}}{d}\right), \qquad (27)$$

and $\mathbf{C}_0$ is defined as

$$\mathbf{C}_0(l) = \begin{cases} -\frac{1}{4}l^5 + \frac{1}{2}l^4 + \frac{5}{8}l^3 - \frac{5}{3}l^2 + 1, & 0 \leq l \leq 1, \\ \frac{1}{12}l^5 - \frac{1}{2}l^4 + \frac{5}{8}l^3 + \frac{5}{3}l^2 - 5l + 4 - \frac{2}{3}l^{-1}, & 1 < l \leq 2, \\ 0, & 2 < l, \end{cases} \qquad (28)$$

$l = \frac{d_{i,j}}{d}$, $d$ is the localization scale, and $d_{i,j}$ is the spatial spherical distance between the $i$th and $j$th grid points, $\boldsymbol{\rho}_{y,k} \in \mathbb{R}^{m_{y,k} \times r}$ is computed together with $\boldsymbol{\rho}_x \in \mathbb{R}^{m_x \times r}$, and $r$ is the selected truncation mode number (Zhang & Tian, 2018), $\mathbf{P}_{x,j}^*$ ($j = 1, \cdots, N$) is a $m_x \times r$ matrix whose every column is the $j$th column of $\mathbf{P}_x$ and $\mathbf{B} \circ \mathbf{C}$ stands for the Schür product of matrices $\mathbf{B}$ and $\mathbf{C}$, which is a matrix whose $(i, j)$ entry is given by $b_{i,j} \times c_{i,j}$. The definition of the notation "$(\cdot <e> \cdot)$" is provided in (Zhang & Tian, 2018).

In particular, the segmented forecast model (7) within each sub-window $[t_{i-1}, t_i]$ (but not the full forecast model $M_{t_0 \to t_k}$) is utilized to update model simulations (via $\mathbf{x}_{b,i-1} + \mathbf{x}_{t_{i-1}}'^{,l}$) over all sub-windows in each iteration. Obviously, such a segmented forecast model (7) facilitates parallel coding. In addition, and unlike the usual ensemble-based approaches employed to only solve the 4DVar problem (Tian et al., 2018), our i4DVar* method for the first time used ensemble simulations not only to solve the optimization problem, but also to formulate the 4DVar*.

**2.2 Big-data driven NLS-i4DVar***

The ensemble simulations (eq. 9) of i4DVar* are identical to those used in NLS-4DVar (Tian et al., 2018). We can thus incorporate the use of "big data" into NLS-i4DVar* by dividing the total ensemble into two parts (Tian & Zhang, 2019): a pre-prepared (historical) big data ensemble (size $N_h$) and a small online ensemble (size $N_o$). The goal of such big-data-driven sampling scheme (see figure 2 and Tian & Zhang, 2019) was to help increase the precision of the ensemble-based background error covariance and composite tangent models in i4DVar*, with significantly decreased computational costs.

To reprepare the historical big data ensemble (size $N_h \gg N_o$), the 4-D big data ensemble is defined by $\mathbf{P}_{x,h}^{4D} = \left( \mathbf{x}_{h,1}^{4D}, \cdots, \mathbf{x}_{h,N_h}^{4D} \right)$, where $\mathbf{x}_{h,j}^{4D} = \left( \mathbf{x}_{h,j}^{t_0}, \cdots, \mathbf{x}_{h,j}^{t_S} \right)$, ($j = 1, \cdots, N_h$), is extracted from the historical forecast simulations (e.g., ensemble simulations started from the initial ensemble states produced by an improved sampling algorithm; Zhang, 2019) over the same-length assimilation window. First, this part of the MPs $\mathbf{P}_{x,i-1}^{h} = \left( \mathbf{x}_{i-1,1}^{h,'}, \cdots, \mathbf{x}_{i-1,N_h}^{h,'} \right)$ and thus

$$\mathbf{P}_{x,h} = \left[ \left( \mathbf{P}_{x,0}^{h} \right)^{\mathrm{T}}, \ \left( \mathbf{P}_{x,1}^{h} \right)^{\mathrm{T}}, \ \cdots, \ \left( \mathbf{P}_{x,N_\tau-1}^{h} \right)^{\mathrm{T}} \right]^{\mathrm{T}} \qquad (29)$$

and their corresponding simulated observation perturbations (OPs) $\mathbf{P}_{y,i-1}^{h,k} = \left( \mathbf{y}_{i-1,1}^{h,k'}, \cdots, \mathbf{y}_{i-1,N_h}^{h,k'} \right)$ and

$$\mathbf{P}_{y,h} = \left[ \left( \mathbf{P}_{y,0}^{h,t_0} \right)^{\mathrm{T}} \ \cdots \ \left( \mathbf{P}_{y,0}^{h,t_1} \right)^{\mathrm{T}} \ \cdots \ \left( \mathbf{P}_{y,N_\tau-1}^{h,t_{N_\tau-1}} \right)^{\mathrm{T}} \ \cdots \ \left( \mathbf{P}_{y,N_\tau-1}^{h,t_{N_\tau}} \right)^{\mathrm{T}} \right]^{\mathrm{T}} \qquad (30)$$

are thus yielded, where $\mathbf{x}_{i-1,j}^{h,'} = \mathbf{x}_{h,j}^{t_{i-1}} - \mathbf{x}_{b,i-1}$ and $\mathbf{y}_{i-1,j}^{h,k'} = L_{k,i-1}\left( \mathbf{x}_{h,j}^{t_{i-1}} \right) - L_{k,i-1}\left( \mathbf{x}_{b,i-1} \right)$. Second, as with the original NLS-4DVar*, a group of online MPs,

$$\mathbf{P}_{x,o} = \left[ \left(\mathbf{P}_{x,0}^o\right)^T, \ \left(\mathbf{P}_{x,1}^o\right)^T, \ \cdots, \ \left(\mathbf{P}_{x,N_\tau-1}^o\right)^T \right]^T \tag{31}$$

(with smaller ensemble size $N_o$) are also needed and their corresponding simulated OPs

$$\mathbf{P}_{y,o} = \left[ \left(\mathbf{P}_{y,0}^{o,t_0}\right)^T \ \cdots \ \left(\mathbf{P}_{y,0}^{o,t_1}\right)^T \ \cdots \ \left(\mathbf{P}_{y,N_\tau-1}^{o,t_{N_\tau-1}}\right)^T \ \cdots \ \left(\mathbf{P}_{y,N_\tau-1}^{o,t_{N_\tau}}\right)^T \right]^T \tag{32}$$

can be obtained through ensemble simulations over the assimilation window $[t_0, t_S]$. Third, we combine the historical big ensemble $\mathbf{P}_{x,h}$ and the online MPs, $\mathbf{P}_{x,o}$, into one set of MPs $\mathbf{P}_x = (\mathbf{P}_{x,h}, \mathbf{P}_{x,o})$ and, similarly, we can obtain the simulated OPs $\mathbf{P}_y = (\mathbf{P}_{y,h}, \mathbf{P}_{y,o})$ Finally, substituting $\mathbf{x}' = \mathbf{P}_x \boldsymbol{\beta}$ and the ensemble background error covariance $\mathbf{Q}_b = \dfrac{(\mathbf{P}_x)(\mathbf{P}_x)^T}{N-1}$ into eq. (3) yields the iterative scheme (23-24).

The total ensemble $\mathbf{P}_x$ is updated by the following modified square root analysis scheme

$$\mathbf{P}_x = \mathbf{P}_x \mathbf{V}_2 \Phi^T \tag{33}$$

See (Tian et al., 2020) for the definitions of $\mathbf{V}_2$ and $\Phi$.

Next, the small online ensemble is updated as

$$\mathbf{P}_{x,o} = \mathbf{P}_x(:,1:N_o) \tag{34}$$

Finally, update the historical big data ensemble partially using the online 4D samples output from the current assimilation cycle as follows

$$\mathbf{x}_{h,j}^{4D} = \mathbf{x}_{h,j+N_o}^{4D}, j = 1, \cdots, N_h - N_o, \tag{35}$$

$$\mathbf{x}_{h,j+N_h-N_o}^{4D} = \mathbf{x}_{o,j}^{4D}, j = 1, \cdots, N_o. \tag{36}$$

For more details please see (Tian and Zhang, 2019). The historical big data ensemble is continually updated through equations (35-36), which realizes its partial flow dependence to a great extent.

**2.3 Forecast model, and observational and evaluation data**

Two-dimensional (2D) shallow-water (SW) equations are used to derive the forecasts of the numerical evaluations (for more details, see Tian et al., 2018). The computational domain comprises $45 \times 45$ grids with $d = 300$ km (i.e. the uniform grid size). The model state vector is represented by height $h$ and the horizontal velocity components $u$ and $v$ at all grid points. The model time step is 360 s (6 min). The "true" initial fields for the evaluation experiments are produced by integrating the perfect 2D SW model with $H_0 = 250$ m at certain initial conditions (eqs.(22-23) in Tian et al. 2021) over 60 h. The background state $\mathbf{x}_b$ is produced using the imperfect model with $H_0 = 0$ m. Thus, $\mathbf{x}_b$ differs significantly from the true state given the two 60-h model integrations at $H_0 = 0$ and $H_0 = 250$ m, respectively.

Observations are available every 3 h (i.e., at 3, 6, 9, and 12 h during each assimilation window of time length 12 h). Each grid has a random observation site yielding 44 × 44 observations at each time point. Observations are generated by adding random noise to the true values at the observation locations, using a simple bilinear interpolation method.

## 3. Results

We evaluated the performance of strongly constrained 4DVar, i4DVar, and i4DVar* using the appropriate ensemble nonlinear least squares-based approaches (i.e., NLS-4DVar, NLS-i4DVar and NLS-i4DVar*, respectively). Remarkably, NLS-i4DVar* required only a very small online ensemble ($N_o = 20$) and a historical "big data" ensemble ($N_h = 40$). All three methods (4DVar, i4DVar and i4DVar*) performed well in the context of the 2D SW model (Tian et al, 2018; Tian and Zhang, 2019) that assumed that the only system-related errors were initial errors, thus producing small root mean square errors (RMSEs; figure 3a,b). Our proposed i4DVar* approach outperformed the other approaches; the RMSEs of i4DVar* were lowest, and those of i4DVar and 4DVar were nearly equivalent at an ensemble size of $N = 60$. However, the latter two approaches exhibited unique strengths and shortcomings. i4DVar performed slightly better and worse than 4DVar in terms of the *height* and *wind* parameters, respectively, as denoted by the RMSEs (figure 3a,b). Notably, i4DVar* used a big-data-driven sampling scheme (Tian and Zhang, 2019) with a small online ensemble size of $N_o = 20$ (figure 3a,b). Thus, i4DVar* achieved superior performance with a relatively small online ensemble ($N_o = 20$) compared to i4DVar and 4DVar using a larger online ensemble ($N = 60$) (figure 3a,b). The spatial distribution of the 4DVar-produced analytical increment $\mathbf{x}'$ (figure 4a) reflected the "true" initial perturbation $\mathbf{x}'_t(0h) = \mathbf{x}_t(t_0) - \mathbf{x}_b$ (subscript $t$ denotes the true state) almost perfectly (figure 4d), which is apparently consistent with the well-accepted "perfect model" assumption of traditional 4DVar. The pattern of the i4DVar-produced integral correction $\mathbf{x}'$ (figure 4b) was also very similar to that of the true perturbation $\mathbf{x}'_t(0h)$ but was of considerably smaller amplitude because the correction $\mathbf{x}'$ was added only at the outset for 4DVar, as opposed to at several selected time steps ($= N_\tau$; corrections $\mathbf{x}'$ were sequentially added along with the model integrations) for i4DVar, which probably explains the differences in $\sim N_\tau$ (=12 in this study) in terms of the $\mathbf{x}'$ amplitudes. The pattern of the i4DVar*-produced integral correction $\mathbf{x}'_{0h}$ at $t_0$ was more similar to that at $\mathbf{x}'_t(0h)$ produced by the 4DVar (figure 4a,c,d), simply because the i4DVar*-produced $\mathbf{x}'_{t_{i-1}}$ ($\mathbf{x}'_{t_0}$) play very similar roles within each sub-window $[t_{i-1}, t_i]$ ($[t_0, t_0 + \tau]$), similar to the 4DVar-produced $\mathbf{x}'$ over the entire assimilation window $[t_0, t_S]$. Obviously, the length of $[t_0, t_0 + \tau]$ is only $\tau$ (= 10 time steps in this study), which is much smaller than $N_\tau \times \tau$ (= $t_S - t_0$).

In the most common situation in which an imperfect model is used ($H_0 = 0$ m), the inferiority of 4DVar was most obvious. First, the performance for $N = 60$ was significantly worse than that for $N = 120$; second, performance was poorer than that of i4DVar at $N = 60$ even when the ensemble size was increased to 120 (figure 3c,d).

This is because the 4DVar analytical increment depends only on the analytical time point; performance will be poor when model error is large, because such error develops at every time point in the assimilation window rather than only initially. In contrast, the i4DVar strategy (sequential correction of error evolution at the opening times of all sub-windows) suppresses the evolution of model error, which explains the superior (compared to 4DVar) performance at small ensemble numbers (figure 3c,d). In addition, the patterns of the i4DVar- and 4DVar-produced $\mathbf{x}'$ values differ greatly from that of $\mathbf{x}'_t(0h)$ (figure 4d,e,f), which is mainly explained by the fact that i4DVar and 4DVar are defined to identify integral corrections when counterbalancing all system-related errors (i.e., not just the initial errors). Similarly, the i4DVar-produced $\mathbf{x}'$ values are of considerably lower amplitude (figure 4f). Our i4DVar$^*$ method yielded substantially lower RMSEs for both *height* and *wind* than i4DVar (figure 3c,d). Importantly, the pattern of the i4DVar$^*$-produced correction $\mathbf{x}'_{0h}$ at $t_0$ remained very similar to that of $\mathbf{x}'_t(0h)$ (figure 4d,g), for the reasons given above. The i4DVar$^*$-produced $\mathbf{x}'_{0h}$ values simultaneously and indiscriminately correcting the initial and model errors in $[t_0, t_0 + \tau]$, thus rendering $\mathbf{x}'_{0h}$ and $\mathbf{x}'_t(0h)$ very similar (figure 4d,g) when the length $\tau$ of $[t_0, t_0 + \tau]$ is relatively short; this also applies to the other sub-windows $[t_{i-1}, t_i]$ (figure 4i–l). Briefly, our i4DVar$^*$ method divides the entire assimilation window into $N_\tau$ small sub-windows and each correction term $\mathbf{x}'_{t_{i-1}}$ simultaneously handles the initial and model errors in $[t_{i-1}, t_i]$ over a relatively short period $\tau$, which dramatically increases the flow-dependency of $\mathbf{x}'_{t_{i-1}}$.

The performance of all three methods is further carefully checked in the first assimilation window (figure 5). It was found that the RMSEs of both 4DVar and 4DVar$^*$ generally change very little throughout the whole first assimilation window. Noticeably, the RMSEs of i4DVar for both *wind* and *height* parameters exhibit a sharp ladder-like downward trend, especially in the first 40 time steps (Tian et al., 2021). Actually, a careful examination shows that the same thing happens in 4DVar$^*$ but with substantially smaller amplitude. This is obviously due to the fact that the "integral" corrections are sequentially added at the opening times of the sub-windows along the integration with model in both i4DVar and i4DVar$^*$. In additional, the RMSEs of the i4DVar$^*$ are uniformly smaller than those of i4DVar over the whole assimilation window (figure 5), which suggests that i4DVar* can easily achieve higher assimilation/forecast accuracy than i4DVar inside/outside of the assimilation window.

i4DVar$^*$ can handle big data, and provides superior performance with a low computational cost when dealing with the large online ensembles encountered in the real world. The incorporation of a segmented forecast model (7) into 4DVar* facilitates parallel coding and reduces computational costs. As shown by Tian and Zhang (2019), our big-data-driven sampling scheme is similar to a hybrid-4DVar method (Clayton et al., 2013), wherein the historical ensemble is analogous to a climatological ensemble (which yields a climatological covariance $\mathbf{B}_c$). The online ensemble reproduces the instantaneous and flow-dependent perturbations, as does its hybrid-4DVar counterpart, probably because inclusion of the climatology-like ensemble of the current big-data-driven 4DVar$^*$ approach ($N_h = 40$ and $N_o = 20$) increases performance slightly compared to fully online sampling ($N_h = 0$ and $N_o = 60$; data not shown). This may change as the dimensions of the forecast model states vary; more research is required on this. The 4D moving sampling strategy (Tian and Feng, 2015) could replace small online ensemble simulations, thereby further reducing the computational cost.

## 4. Discussion and conclusions

Currently, model errors are corrected using only strongly or weakly constrained methods, which can be distinguished according to whether the 4DVar solution must exactly or approximately satisfy the forecast model (Trémolet, 2006, 2007). Thorough analysis illustrated that the two approaches balance the degrees of freedom and error rate differently. The strongly constrained 4DVar restricts the optimized variable via analysis of the increment $\mathbf{x}'$ at $t_0$, thereby reducing the degrees of freedom ($N_v \times n_{\dim}$). A weakly constrained 4DVar aims to distinguish the initial and model errors and correct them separately at higher degrees of freedom (($S+1) \times N_v \times n_{\dim}$). The optimization problem remains underdetermined unless additional information is added. Various simplifications have been proposed to deal with model error, all of which add information and then apply different constrained 4DVar approaches.

The hidden mechanism of the strongly constrained 4DVar that corrects both initial and model errors was first revealed by Tian et al. (2021), who formulated an integral-correcting 4DVar (i4DVar) by extending the analysis from the initial time to other times. However, i4DVar integral correction is simply an averaged correction term that suppresses initial and model error evolution at selected time steps (Tian et al., 2021) by dividing the whole assimilation window $[t_0, t_S]$ into $N_\tau$ sub-windows. This averaged/uniform correction term is flow-independent over the entire assimilation window. To address this issue, flow-dependent correction terms $\mathbf{x}'_k$ were introduced to increase the degrees of freedom ($N_\tau \times N_v \times n_{\dim}$). However, unlike typical weakly constrained 4DVar approaches, i4DVar* employs ensemble simulations (eq. 9) to construct a solution space of $\Omega(\mathbf{P}_x)$ for the correction term $\mathbf{x}'$, thereby rendering the i4DVar* problem solvable even if there are many degrees of freedom ($= N_\tau \times N_v \times n_{\dim}$). This is the first study to use ensemble simulations not only to solve the 4DVar problem, but also to formulate the approach. Ensemble simulations have commonly been employed to approximate background error covariance and constitute adjoint-free approaches to 4DVar (i.e., the so-called 4DEnVar methods; see Tian et al., 2018). Encouragingly, numerical evaluations show that our i4DVar* method is easy to program, exhibits high-level precision during DA, and is computationally efficient, especially when dealing with big data. Compared to traditional 4DVar and i4DVar methods, 4DVar* exhibits superior performance, and simple and inexpensive to implement.

**Conflict of Interest**

The authors declare no conflicts of interest relevant to this study.

**Data availability statement**

 The data used in this paper are all freely available

 (https://doi.org/10.5281/zenodo.7325269).

**Acknowledgments**

   This work was jointly supported by the Second Tibetan Plateau Scientific Expedition and Research Program (2022QZKK0101) and the National Natural Science Foundation of China (Grant Nos. 41575100, 42105150).

**Table 1** Comparisons between s4DVar, w4DVar, i4DVar and i4DVar*

|         | Degrees of freedom | Model constraints | Dis. initial and model errors |
|---------|---------------------|-------------------|-------------------------------|
| s4DVar  | $N_v \times n_{dim}$ | Strongly | No |
| w4DVar  | $(S+1) \times N_v \times n_{dim}$ | Weakly | Yes |
| i4DVar  | $N_v \times n_{dim}$ | Weakly | No |
| i4DVar* | $N_\tau \times N_v \times n_{dim}$ | Weakly | No |

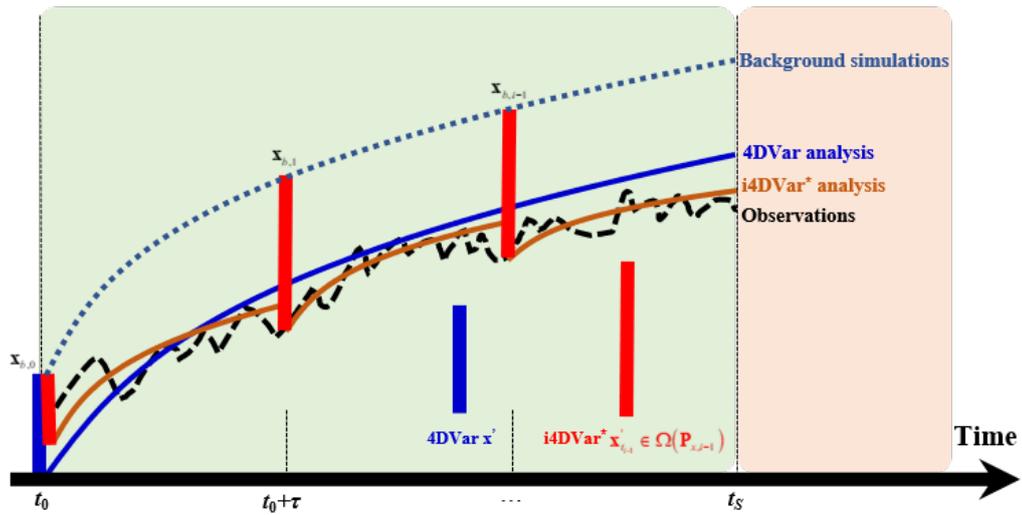

**Figure 1**. Schematic diagram of i4DVar* vs. 4DVar

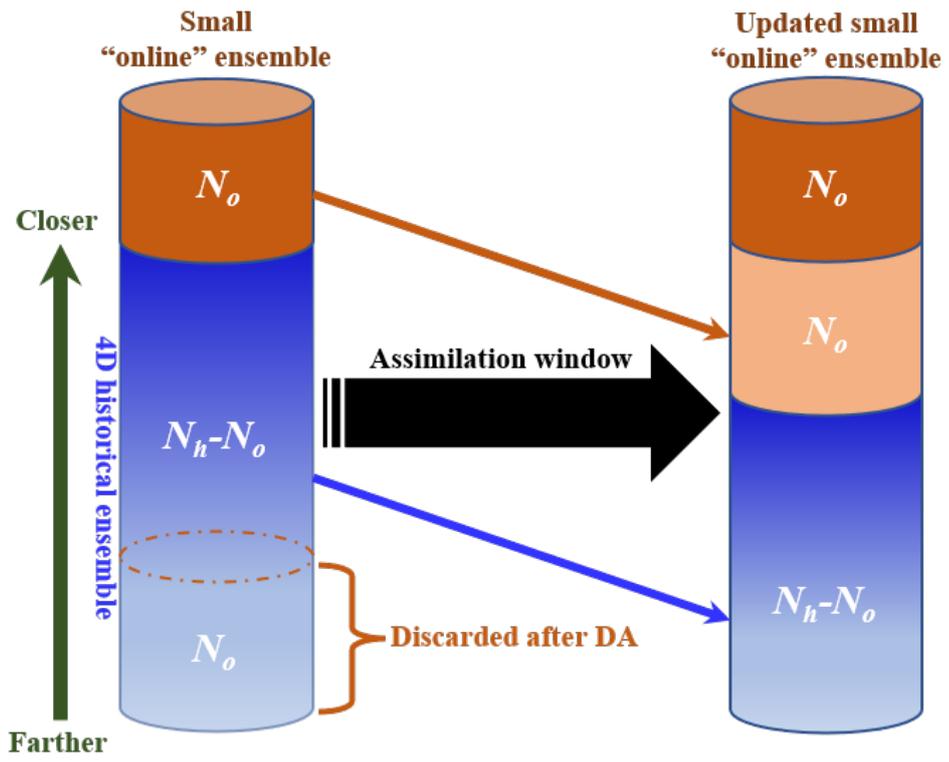

**Figure 2**. Big-data-driven sampling scheme

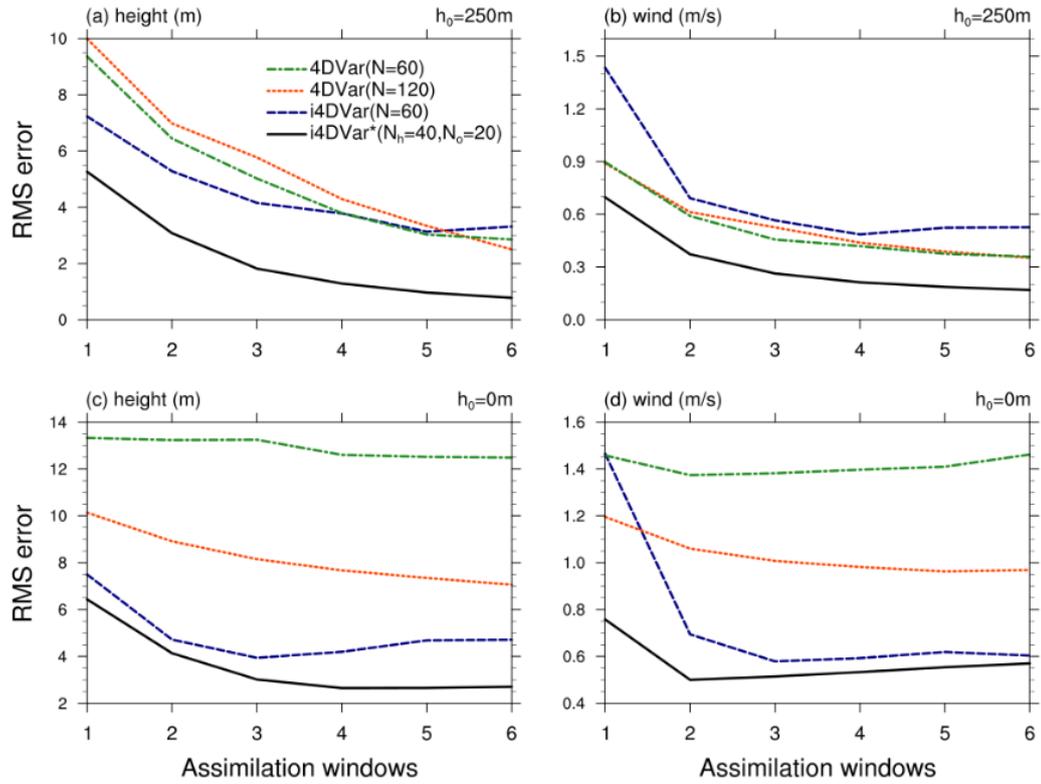

**Figure 3**. Time series of spatially and temporally averaged root mean square errors (RMSEs) of height (left column) and wind (right column) for 4DVar ($N$ = 60, 120; N is the ensemble number), i4DVar ($N$ = 60) methods and i4DVar*($N_h$=40 and $N_o$=20) under the scenarios of the (a and b) perfect model ($h_0$ = 250 m), (c and d) imperfect model ($h_0$ = 0 m), respectively.

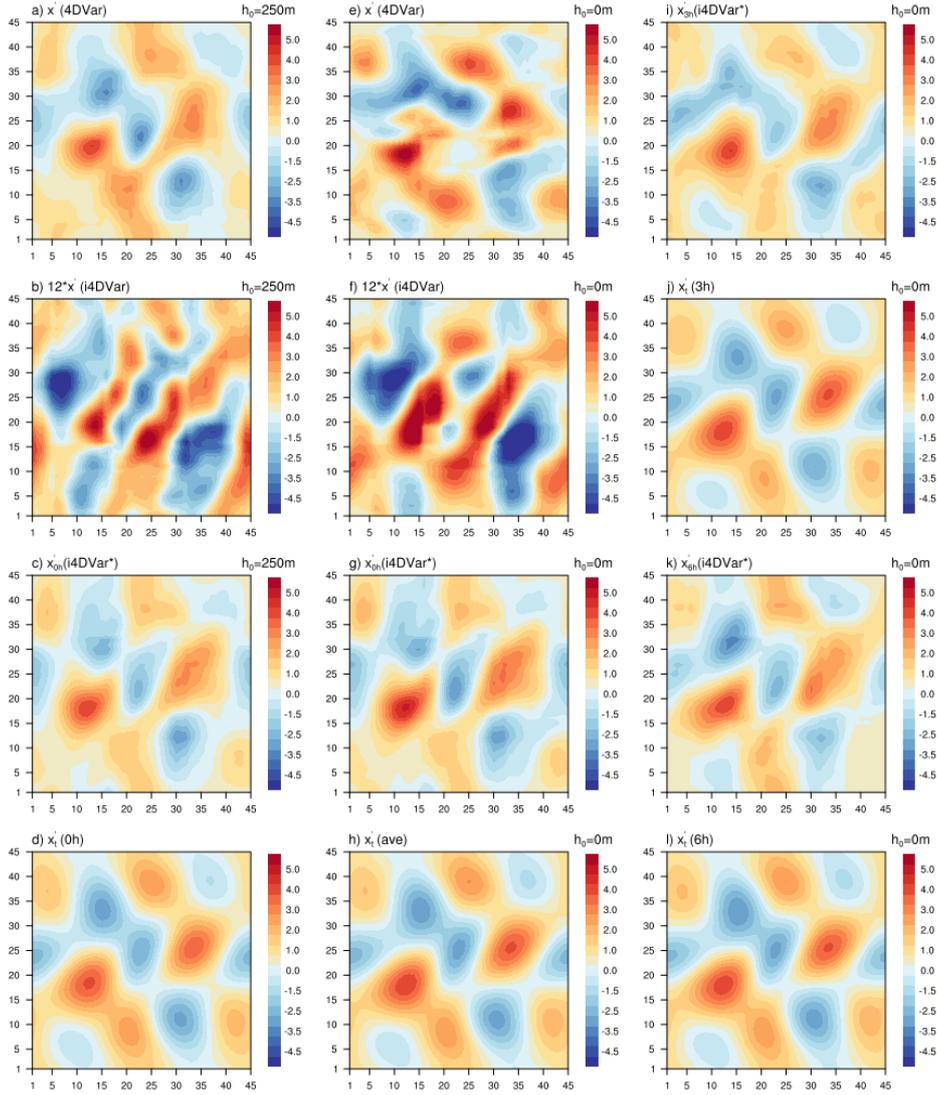

**Figure 4**. Spatial distribution of the correction-term/increment (the horizontal velocity component *u*) from the 4DVar, i4DVar, i4DVar* at $t_0$ and the initial "true"/averaged perturbations under the scenarios of the (a–d) perfect model ($h_0$ = 250 m), (e–h) imperfect model ($h_0$ = 0 m); and (i-l) from the i4DVar* and the "true" perturbations at 3h and 6h under the scenario of the imperfect model ($h_0$ = 0 m), in the first assimilation window, respectively.

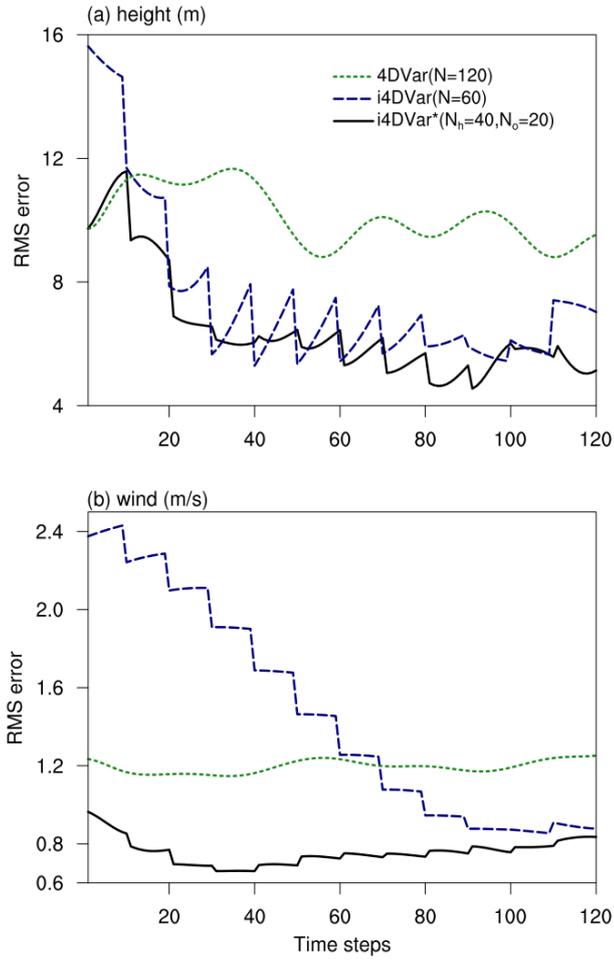

**Figure 5**. Time series of spatially averaged root mean square errors (RMSEs) of (a) height and (b) wind for the 4DVar ($N$ =120), i4DVar ($N$ = 60) and i4DVar$^*$($N_h$=40 and $N_o$=20) methods in the first assimilation window under the scenario of the imperfect model ($h_0$ = 0 m), respectively.